\documentclass[runningheads]{llncs}
\usepackage{makeidx,epsfig}
\usepackage{graphicx}
\usepackage{setspace,graphicx}

\usepackage{booktabs}
\usepackage{multirow}

\usepackage[margin=1.5in,footskip=0.5in]{geometry}

\begin{document}
\title{MAC Protocols for Wireless Mesh Networks with Multi-beam Antennas: A Survey\thanks{Published in Proceeding of  Future of Information and Communication Conference (FICC) 2019       DOI: https://doi.org/10.1007/978-3-030-12388-8\_9}}

\author{Gang Wang, Yanyuan Qin}

\authorrunning{Gang Wang et al.} 
\institute{Dept. of Computer Science and Engineering, University of Connecticut \\
\email{Email: g.wang.china86@gmail.com}}

\maketitle

\begin{abstract}
Multi-beam antenna technologies have provided lots of promising solutions to many current challenges faced in wireless mesh networks. The antenna can establish several beamformings simultaneously and initiate concurrent transmissions or receptions using multiple beams, thereby increasing the overall throughput of the network transmission. Multi-beam antenna has the ability to increase the spatial reuse, extend the transmission range, improve the transmission reliability, as well as save the power consumption. Traditional Medium Access Control (MAC) protocols for wireless network largely relied on the IEEE 802.11 Distributed Coordination Function(DCF) mechanism, however, IEEE 802.11 DCF cannot take the advantages of these unique capabilities provided by multi-beam antennas. This paper surveys the MAC protocols for wireless mesh networks with multi-beam antennas. The paper first discusses some basic information in designing multi-beam antenna system and MAC protocols, and then presents the main challenges for the MAC protocols in wireless mesh networks compared with the traditional MAC protocols. A qualitative comparison of the existing MAC protocols is provided to highlight their novel features, which provides a reference for designing the new MAC protocols. To provide some insights on future research, several open issues of MAC protocols are discussed for wireless mesh networks using multi-beam antennas.
\end{abstract}



\section{Introduction}
Due to a growing popularity of wireless local access, there exists a high demand to improve network throughput and enhance energy efficiency in data transmission between terminal devices (e.g., mobile phones) and access points (or base stations). However, wireless local networks mainly focus on the single-hop transmission. With the help of explosive implementations of wireless network in practical, it has sparked the idea of Wireless Mesh Networks (WMN)~\cite{00_WMN}, which can potentially improve the overall network capacity, enlarge the network coverage, and facilitate the network's auto-configuration. WMN networks typically ogrnaize the communication nodes in a mesh topology, which is similar to a wireless ad-hoc network. In a WMN, it typically has three components:  terminals, routers (or switches) and gateways~\cite{00_WMN1}. Compared to traditional antennas,  multi-beam antennas, alternatively called smart antennas, provide several advantages, e.g., a higher antenna gain,  longer transmission range, better spatial reuse, and much lower interference~\cite{07}. Thus, introducing multi-beam antennas into wireless LAN can improve the overall performance of a network, especially for the wireless mesh networks. For example, WMN and its applications can be largely used in a harsh working condition or a disaster relief environment to provide special services~\cite{0102}. The advantages of multi-beam antenna on WMN have attracted the researchers from both the academy and the industry, which result in rapid commercialization with lots of standardization efforts~\cite{01}. 

An omni-directional antenna typically spreads its wireless electromagnetic energy over a much larger area, however,  only a small part can be actually received by a intended receiver, thus potentially wasting lots of transmission energy. Besides, most omni-directional antennas have some common issues, e.g., delay spread, multipath fading and co-channel interference (CCI)~\cite{0318}. Due to more efficient signal processing algorithm and low-cast computing capacity, the beamforming antennas are available to wireless communication systems~\cite{0304}~\cite{03}. Antennas with beamforming technology use arrays of simple smart antennas  (MBA) to improve the transmission efficiency.  The multi-beam antennas can actively control the temporal paces, by using the DSP (Digital Signal Processing) units, to enhance the radiating electromagnetic energy. 

Wireless LAN typically equips with a finite set of (mobile) stations and the Access Point (AP).  Compared to mobile stations, access point is much more powerful as well as less physically constrained, and both can be viewed as a Full Function Unit (FFU). To increase the network throughput and exploit the spatial reuse, access point typically equips with several smart antennas~\cite{02}. For existing multi-beam antennas, it can be roughly categorized into three classes: adaptive array antennas, switched multi-beam antennas and multiple-input-multiple-output(MIMO) links~\cite{0216}. We will discuss these pros and cons later. Due to simple and commercially available, the  switched multi-beam antennas have been deployed  in lots of practical applications~\cite{07}.

To achieve a superior capability for smart antennas, we can appropriately design its upper layer protocols in protocols stack, such as MAC layer protocols. Compared with a traditional MAC protocol, it sill has several other design challenges in MAC protocol of smart antennas.  In traditional network, most wireless nodes equipped with omnidirectional antennas, and its protocols are simple. For example, we do not need to consider the interference among the underlying beams. If without an appropriate control, this interference may deteriorate the overall communication performance, which can be even below the performance of omnidirectional antennas~\cite{0110}. Thus, it is necessary to design appropriate protocols which are suitable for multi-beam antennas in WMNs. Lot of great efforts have already achieved for MAC layer protocols, and this paper provide a deep survey in this area.

The rest paper is organized as follows. In section~\ref{Sec:Basic}, we introduces the background information of multi-beam antennas. Section~\ref{sec:challenges} describes current design challenges in beamforming antennas. Section~\ref{Sec:Class} presents MAC protocols classification. Section~\ref{Sec:claasic} surveys classic MAC protocols for WMNs with multi-beam antennas. Section~\ref{Sec:dis} discusses the problems with current design and predication.  Section~\ref{Sec:Concl} concludes this paper.

\section{Basics of Multi-Beam Antennas}
\label{Sec:Basic}
In this section, we provide some concise knowledge about the multi-beam antennas, on MAC protocols, in wireless mesh networks.

\subsection{Multi-Beam Smart Antennas}
The wireless mesh networks typically extend the infrastructure-based single-hop wireless network~\cite{00_WMN}. Initially, almost all wireless architectures assume the use of omni-directional communication in a wireless mesh network, which causes poor spatial reuse in multi-hop networks, adversely affecting the network capacity~\cite{0502}. The transmission capacity can be enhanced considerably by using smart antennas, given their better spatial reuse~\cite{0503}. Recent researches have investigated the applicability of multi-beam antennas in wireless mesh networks.

Multi-beam smart antenna, shortly named as multi-beam antenna, is referred to a multiple beam antenna array. This antenna array can simultaneously transmit (or receive) multiple packets on different beams using the same channel, thus it can substantially improve the single-hop throughput and the overall throughput~\cite{05}. However, simultaneous transmission (or reception) at a same node typically requires the corresponding antennas equipped with both spatial multiplexing and demultiplexing capability, which has been called Space Division Multiple Access(SDMA) in literature~\cite{0304}~\cite{0413}. 

Before discussing the multi-beam antenna, it is necessary to discuss the beamforming antennas for easy understanding. Any radio-based antennas can provide its primary function that couples electromagnetic energy from one medium to another of the same type. For omni-directional antennas, their simple dipole antennas can be used to radiate/receive energy equally to/from all directions. For the directional antenna, another type smart antenna, it can be able to radiate/receive energy to/from one specific direction more than the others~\cite{01}. 

To quantify Quality of Service(QoS) of the antennas, one of the most important characteristics is the antenna's gain, which can be used to measure the QoS. Usually, the gain is used in the directional antennas, which indicates the relative power in a certain direction compared to omni-directional antennas, the gain is often measured in \textit{dBi}. Specifically, the gain of an omni-directional antenna equals 0 \textit{dBi}. For the reciprocal antennas with the characteristics of transmission and reception, the gains can be further separated into transmission gains and reception gains. However, it is difficult to obtain the exact gain values due to the properties of wireless signals,  and the gain values in all directions of space can be represented by the antenna radiation pattern. A directional antenna pattern usually consists of a high gain main lobe and several gain side and back lobes~\cite{01}.

\begin{figure}
  \centering
  \includegraphics[width=8cm]{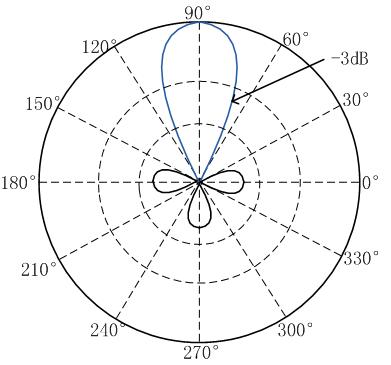}
  \caption{Antenna Radiation Pattern with a main lobe pointing 90$^\circ$ and side lobes with small gains}
  \label{fig:0101}
\end{figure}

Fig.~\ref{fig:0101} shows an example of an antenna radiation pattern with the main lobe pointing to $90^{\circ}$ and side lobes with smaller gains. The axis of the main lobe is known as the boresight of the antenna, which also lies along the peak gain, the maximum gain over all directions. The beam width formally refers to the angle subtended by the directions on either side of boresight, are $3dBi$ less in gain. Ideally, the directional antennas are assumed to have an ideal antenna pattern in which the gain is a constant value in the main lobe and zero outside of the main lobe. However, it is not practical in real design applications for ideal antennas due to the existing of the interference, especially for the multi-beam antennas.

The smart antenna usually equipped with antenna arrays with physical separation in terms of a fraction of the wavelength. It can produce a specific antenna radiation pattern. The overall radiation pattern of an antenna array is determined by several important parameters, such as the number of the single antenna elements (e.g., dipoles), the space among the elements, the geometrical configuration of the array as well as the amplitude and phase of the applied signal to each element~\cite{01}.

Beamforming antenna is a type of smart antenna which includes a Multiple Input Multiple Output(MIMO) control system~\cite{0114}. This system combines the antenna array with Digital Signal Processing (DSP) techniques to allow the transmission and reception for the antenna elements. Usually, the beamforming antenna employs the sophisticated antenna array schedule and control algorithms to automatically and adaptively control the overall radiation patterns of the antennas. In this paper, the beamforming antennas can be simply and interchangeably referred to as directional antennas. 

\begin{figure}
  \centering
  \includegraphics[width=8cm]{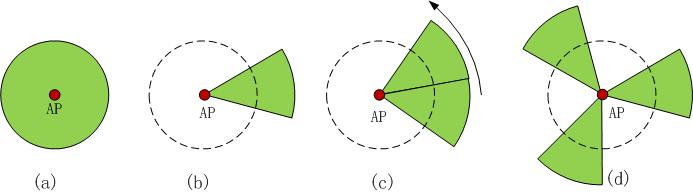}
  \caption{The coverage range of different transmission modes}
  \label{fig:0110}
\end{figure}

Fig.~\ref{fig:0110} shows the coverage range of different transmission modes: (a) Omni-directional mode. (b) Uni-directional Mode. (c) Multi-directional Sequential Mode. (d) Multi-directional Concurrent Mode.

Compared to the traditional omni-directional antennas~\cite{0115}, there exists several potential benefits and numerous advantages.

\begin{enumerate}
\item Significantly reducing interference: Due to the radiated energy in the direction of the intended receiver of directional antennas, the transmission (or reception) does not interfere too much with neighboring nodes residing in other directions.
\item Increased Signal-to-Noise Ratio(SNR): With the same transmit power, the gain of beamforming antennas focuses more energy in the intended direction to increase SNR, moreover, the link quality and transmission rate.
\item Extended communication ranges: In wireless mesh networks, the extension of communication range may lead to fewer-hops routes and consequently reduce the end-to-end delay~\cite{0119}, also may improve the connectivity of the network~\cite{0121}.
\item More energy efficient communication~\cite{0122}.
\item More secure wireless communication: The beamforming antenna can reduce the risks of eavesdropping and jamming to provide more secure communication~\cite{0124}~\cite{0125}.
\item Location estimation~\cite{0126} and efficient broadcasting~\cite{0127}
\end{enumerate}

\subsection{Medium Access Control(MAC)}
To take the benefits of multi-beam smart antennas, the link layer needs to be properly designed for providing the service to the network layer. IEEE 802.* standards provide the corresponding protocols for the wireless LANs and Ad Hoc networks. One of the goals of the MAC protocol is to set the rules in order to enable efficient and fair sharing of the common wireless channel~\cite{0128}~\cite{0129}. MAC protocols for wireless networks could be classified into two major categories: contention-free MAC and contention-based MAC~\cite{0130}. 

Contention-free MAC is largely based on the controlled access in which the channels are, according to the predetermined schedule, allocated to each node. Contention-based MAC protocol usually is implemented through the random access, which the nodes in it compete to access the shared medium. When there happens a conflict, the distributed conflict resolution algorithm is used to resolve the collision.

IEEE 802.11, the de facto standard for medium access control in wireless networks, is typically designed for the omni-directional antennas in wireless communication~\cite{0504}. IEEE 802.11 standard provides one mandatory channel access function DCF (Distributed Coordination Function) and one optional channel access function PCF (Point Coordination Function). In the construction of these two types, the difference is that PCF is centralized, while DCF is fully distributed. Here we describe the general design of DCF in MAC layer, then we will discuss the design challenges for multi-beam antenna in MAC layer.

IEEE 802.11 DCF is a carrier sense based MAC protocol, which employs CSMA/CA (Carrier Sense Multiple Access with Collision Detection) mechanisms at the MAC layer. The CSMA/CA provides contention-based single-channel access to APs in the network.

\begin{figure}
  \centering
  \includegraphics[width=8cm]{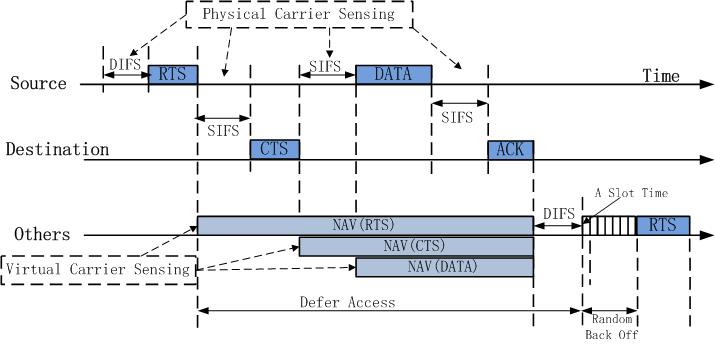}
  \caption{Basic Operation of IEEE 802.11 DCF}
  \label{fig:0103}
\end{figure}

In CSMA/CA mechanism, as shown in Fig.~\ref{fig:0103}, when a node wishes to transmit the data, it first performs physical carrier sensing before initiating transmission, which is the CSMA part of the CSMA/CA protocol. There exists the case that two nodes, each is outside the carrier sensing range, are trying to transmit data with a common node. In this case, a collision occurs at the receiving node. To avoid the collision happening, collision avoidance, the CA part of the protocol, is implemented by a handshaking mechanism before data transmission~\cite{0132}. For the handshaking mechanism, if the sender senses the channel at the receiver site as idle for a Short Interframe Spacing (SIFS) period, the sender transmits a short Request-To-Send (RTS) packet to the intended receiver and the receiver in turn responses with a short Clear-To-Send (CTS) packet. Both RTS and CTS packets contain the proposed duration of the transmission. Above process is the handshaking mechanism. If the channel is still sensed idle, the sender waits for a DCF Interframe Spacing(DIFS) period before sending its packet. If the channel is busy with physical sensing after this waiting period, then the node chooses a random backoff duration from the set of [0, \textit{CW}]. This duration is determined by the contention windows (\textit{CW}). The CW is an integer between \textit{$CW_{min}$} and \textit{$CW_{max}$}, where \textit{$CW_{min}$} and \textit{$CW_{max}$} depend on physical layer characteristics. Initially, the \textit{CW} is set to the value of \textit{$CW_{min}$}. The node decreases the backoff timer by one after each idle slot time. When the timer equals to zero and the channel which the sender sensed is idle, the sender can transmit its packet. If any activity is detected on the channel during the backoff period, the node freezes its backoff timer and waits until the channel is idle again. In the backoff algorithm, when the backoff timer expires, the sender doubles its \textit{CW}, chooses a new backoff interval and tries retransmission again. The \textit{CW} is doubled on each collision until reaching a maximum threshold, \textit{$CW_{max}$}. Also, the number of retransmission attempts is limited by the threshold after which the packet is discarded. In the case of the CTS or ACK packet not received back, the sender assumes that a collision has occurred with some other transmission and it invokes the binary exponential backoff algorithm as mentioned above. When the transmission successfully transmitted between the sender and receiver, the contention window \textit{CW} is initialed to its minimum value for the next transmission. 

The above CSMA/CA mechanism mitigates the probability of two nodes transmitting data at the same time. The random deferral by each node can ensure fair channel award in the long term and is given by 

$Random\_Backoff = Random() \times Slottime$,

where \textit{Random()} returns a pseudo-random integer from a uniform distribution over the interval[0, \textit{CW}] and \textit{Slottime} is related to the corresponding physical layer characteristics.

From the design of IEEE 802.11 standard, it implicitly assumes an omni-directional antenna at the physical layer. However, when using the multi-beam smart antennas, IEEE 802.11 standards do not work properly.

\section{MAC Design Challenges for Beamforming Antennas}
\label{sec:challenges}
Two types of multi-beam antennas have been extensively accepted and studied in current literature: one is based on the fixed beam directional antenna and the other is based on adaptive arrays. Adaptive-array based smart antenna may work better in a multi-path rich environment, however, it is more complex to design the transceiver and the corresponding MAC protocols. In this section, we briefly present the major challenges when designing a MAC protocol in beamforming domain.

\subsection{Beam-synchronization constraint}
In the multi-beam antenna, the cooperation among the beams is required so that the AP can work correctly with each beam. To prevent the co-site interference issue, all sections at the AP must be in a signle mode, either the reception mode or the transmission mode~\cite{07}. This needs a sophisticated scheduling scheme to cope with the synchronization between different modes.

\subsection{Hidden terminal problem}
\begin{figure}
  \centering
  \includegraphics[width=8cm]{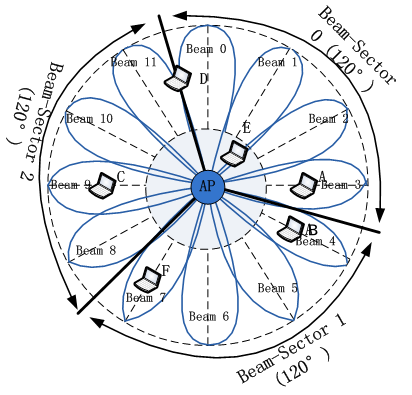}
  \caption{An example of sectorized multi-beam antenna system. Number the beams and sectors in a clockwise direction.}
  \label{fig:0201}
\end{figure}

Hidden terminal problem also is common in a traditional network; it occurrs when two nodes attempt to communicate with a common node, however, both of them are outside of their carrier sensing range during CSMA. For this problem, one solution in traditional network is to implement a RTS/CTS handshaking mechanism before real data transmission, thus avoiding the occurrence of collision~\cite{0132}. For example, as shown in Fig.~\ref{fig:0201}, we assume that a station $F$ wants to communicate with an AP while this AP is communicating with either station E or station B. Station F, without sensing the signal from either station E or station B, infers that the transmission media is free, and then sends its data to this AP, which eventually causes a collided data.

In a multi-beam antenna, new hidden terminal problems can present. For example,  when a potential interferer could not receive the exchange information of RTS/CTS, due to its antenna orientation to others during the handshake, and then initiates a data transmission this may cause a transmission collision. Two new types of hidden terminal problems are presented~\cite{0144}:

Hidden Terminal Due to Asymmetry in Gain

Usually, in the multi-beam antennas, it uses the beamforming techniques to enhance the gains to transmit the data. The gain in a omnidirectional mode is much smaller than the gain in a beamformed scenario. If an idle node is omni-directionally listening to the medium, this node may not unaware of some ongoing transmission that could be affected with its directional transmission~\cite{01}. Thus, this case causes a collision.

Hidden Terminal Due to Unheard RTS/CTS

The loss of the channel state information during beamforming causes this type of hidden terminal problem. When a specific node is engaging in one beamforming communication, it might appear to be deaf to other directions, therefore, important control information may be lost during that time period. When a neighboring node fails to receive this exchanged channel reservation packets  between a transmitter-receiver pair, such as RTS and CTS,  it will cause the hidden terminal problem. In this case, the receiver may become unaware of a imminent communication between that transmitter-receiver pair and consequentally could later initiate another transmission which results the data collision~\cite{01}.

The hidden terminal problem is much more significant in WMN, which needs a more complex MAC protocol to synchronize the exchanged transmission information between the neighbor nodes, especially in the case of dealing with mobility. For traditional access point AP with omni-directional antennas, the mobility may not be a big problem. In a WMN with multi-beam antennas,  it may arise a non-negligible problem for AP node, in which each node is physically associated with a beam-sector. The medium access control, especially the download medium access control, will be highly affected when a mobile node moves from one beam-sector to another. It needs an efficient location updating algorithm to keep the freshness of the location information with reasonable overhead, which increases the complexity of the wireless mesh network~\cite{07}.

\subsection{Deafness}
One of the aims using multi-beamforming antenna is to exploit the spatial reusability, deafness maybe occur and is by far one of the most critical challenges in a wireless network, which was first identified and defined in the basic directional MAC protocol~\cite{0133}~\cite{0141}~\cite{0142}~\cite{0143}. For example, when a transmitter tries to communicate with a receiver, however, this receiver might in a beamforming toward another direction that is away from the transmitter,  this will cause the trying transmission process failed. Different from the characteristics of omni-directional antennas, the intended receiver in the beamforming antenna is unable to receive the transmitter's signal. Thus, it appears deaf to the transmitter.

Fig.~\ref{fig:0201} shows the deafness problem, also known as receiver blocking problem. Assume that the station B (in sector 1) intends to send its data to a access point AP, meanwhile, this AP is sending data to station A (in sector 0). Without hearing this beamformed transmission from the AP to station A, station B assumes that the media which station B will be used is free at the end of its back-off, and then sends transmission data to the AP. Since the AP is multi-beamformed in this case, it needs the beam-synchronization to constrain before receiving or sending data and is deaf to station B's transmission, as the result the AP is unable to receive the data sent from station B. Without getting response, such as CTS, from the AP, station B typically considers this kind of failure as an indication of collision and reacts accordingly. Thus, the station B involves the binary exponential backoff algorithm before attempting retransmissions. Station B may keep sending data until its retry limit is reached. These unnecessary retransmissions reduce the network capacity and lead to significant bandwidth waste. Also, the exponential increase in the backoff contention windows results in channel underutilization. In this case, it may make matters worse, the transmission of station B may corrupt the reception of station A from the AP since station A is located close to station B.

The consequences incurred by deafness problem may be even worse, and this  may potentially leads to a short-term unfairness between the flows that share a common receiver. Furthermore,  if the involved transmitter has multiple packets to send and constantly transmits its data by choosing a backoff interval from its minimum contention windows, this might caues a deadlock eventually. Moreover, a chain of deafness is possible when every station attempting to communicate with a deaf station becomes itself deaf (self-deaf) to another station~\cite{0143}.

\subsection{Beam-overlapping}
Another aim of multi-beam anatenna is to improve the spatial utilization. Multi-beamforming technique implements multiple beamformed beams in the smart antenna, there must exist the interference between different beams. Due to the physical imperfection of beamforming antennas, there generally exists a small portion of the beam-overlapping area for two adjacent beams, as showing in Fig.~\ref{fig:0201}. If a station lies in the beam-overlapping area, this station can hear data transmissions from multiple sectors, in turn, multiple sectors in this access point AP can also hear data transmission from that station. For example, in Fig.~\ref{fig:0201}, assume that stations C and D are simultaneously sending data to the AP. Since both sector 0 and sector 2 can hear the signal from station D, sector 2 will receive a collided data from stations C and D~\cite{02}. This is not what the multi-beam wants.

Moreover, the beam-overlapping problem can cause the back/side-lobe problem. As known that beamforming technique usually has high-gain than that of traditional antennas, even though the APs are equipped with multiple high-gain narrow-beam directional antennas, however, the negative effects of back/side-lobe problem introduced by interference cannot be totally ignored. For example, in Fig~\ref{fig:0201}, when station E intends to send data to the AP, all sectors may receive this signal from station E since it is so close to the AP and falls in the back-lobe or side-lobe of many other beams~\cite{02}.

Also, the beam-overlapping problem is much more serious in the multi-path rich environment. One station, e.g., station C in Fig~\ref{fig:0201}, in the range of AP, may hear  the ongoing transmissions from all sectors, in turn, all other sectors at the AP can also hear the transmission from station C. This is the multi-path rich problem. It potentially degrades the omni-directional communication,  thus, no spatial reuse can be exploited.

\subsection{Unnecessary defer}
For multi-beam antenna, two kinds of unnecessary defer issues exist: one is caused by the rule of CSMA/CA, the other is the Head-of-Line (HoL) blocking problem in beamforming MAC protocols which was first identified in~\cite{0145}. 

For omni-directional antenna, First-In-First-Out (FIFO) is popular in queueing policy to buffer the received signals and this policy works great in an omni-directional antennas due to using the same transmission medium for all outstanding packets. When the medium is a busy state, it would not have any  packets to be transmitted. But, for multi-beamforming antennas, since the transmission medium is spatially divided, and some directions may be available for transmission, while others may not. This is common case, since each beam in multi-bean antenna is independent, and has its own communication ranges. In case that a packet is on the top of its queuing, if using of FIFO scheme, then the transmission is destined to a busy station, which further blocks all the subsequent transmission even though some of them are available to be transmitted. Moreover, HoL blocking issues maybe aggravate in the case that the top packet goes into deadlock (a round of failed retransmissions)~\cite{0146}.
  
Due to the geographically close for each beam in multi-beam antennas, or even existing beam-overlapping, one station may hear another station's transmission, and maybe subdues to transmit the data. For example, in Fig.~\ref{fig:0201}, assume that one of stations in sector 1, say station B, intends to send data to the AP, meanwhile, the AP is receiving data from station A in sector 0. We know that different sectors in the same AP can simultaneously transmit the data to the AP. It is clear that station A and B can simultaneously transmit their respective data to the AP because of the different sectors which station A and station B located in. However, since two adjacent stations, say station A and B, are geographically close to each other, station B can hear station A's signal, even lying in different sectors, and will keep silent according to the rule of CSAM/CA, thus causing the throughput down.

\subsection{Miss-hit}

Miss-hit problem occurs when the AP directs wrong beams for downlink transmission because of station movement. The majority of wireless communication relies on Direction of Arrival (DOA) principle~\cite{0214}. However, the stations within the wireless mesh network are likely to move frequently, which makes the communication even more challenging. Based on DOA estimation, packets are sent from one station to an AP with smart antenna. AP will use these packets to identify the beam or beams (with beam-overlapping or back/side-lobe) from the sender. When station in the mesh network moves, the AP may cache the wrong beam information, which should be updated/corrected based on new location.

\section{MAC Protocols Classification}
\label{Sec:Class}
Since IEEE 802.11 DCF-based WLANs have been widely deployed, the question is how an IEEE 802.11 node transmits and receives data to/from a multi-beam access point with the above-mentioned challenges. The problem of designing an efficient MAC protocol for wireless mesh networks with multi-beam antennas has been of a great interest in recent years. Generally, the MAC protocols in literature can be broadly classified into random access protocols and synchronized access protocols. 

Random access protocols allow the stations lying in different sectors to access the shared medium, say access point AP, randomly through contention with each other. This type of protocol typically employs CSAM/CA to avoid collision among different beams. Synchronized access protocols allow the stations to access the medium based on the predetermined schedule which can be achieved through local and/or global synchronization. Random access protocols can be further classified into sub-categories according to the techniques used to deal with the MAC challenges mentioned above section. One sub-category of random access protocols relies solely on the control packets, such as RTS/CTS, to avoid the collision. The other sub-category employs busy tones that are usually transmitted on a dedicated control channel~\cite{01}.

Synchronized access protocols require some sort of synchronization between the stations to coordinate conflict-free transmissions to occur simultaneously. All beams in multi-beam antenna at the access point AP can be either in the receiving mode or the transmission mode due to the well-known co-site interference problem~\cite{csi}, assuming all the beams operate in the same frequency band. This cause that it is hard to achieve since beam-synchronization (for either reception or transmission) requires to facilitate multiple parallel transmissions. It is a common discipline to allow only one node to transmit at a time. However, in this type of protocol, the time period is usually divided into frames and each frame consists of sub-frames which are simply a group of time slots~\cite{01}. The sub-frames are further divided into two types: one sub-frame is to perform the schedule for channel contention; the rest of sub-frames is to transmit the scheduled contention-free data~\cite{0177}~\cite{0178}~\cite{0180}. Generally, it is difficult to achieve the global synchronization in multi-hop wireless mesh networks, most protocols in literature choose to relay on local synchronization among neighboring stations~\cite{0183}~\cite{0149}.

There are other types classification regarding to the multi-beam antennas in wireless mesh network, according to different specifications. Table~\ref{Tab:class} shows the different classifications. Note that these classifications are not independent of each other and one MAC protocol may belong to more than one classes.

\begin{table}
\begin{center}
\caption{Taxonomy of MAC protocols for wireless mesh network with antennas}
\begin{tabular}{ |c|c| } 
\hline
 & Classifications \\
\hline
\multirow{2}{7em}{Antenna Capabilities} & Switched-beam antennas  \\
& Steered-beam antennas(Adaptive antennas) \\ 
\hline
\multirow{2}{7em}{Communication Range} & Omni-directional Antennas \\ 
& Directional Antennas \\ 
\hline
\multirow{2}{7em}{Channel Number} & Single Channel \\ 
& Multiple Channel \\ 
\hline
\multirow{2}{7em}{Access Medium} & Random Access \\ 
& Synchronized Access \\ 
\hline
\end{tabular}
\label{Tab:class}
\end{center}
\end{table}
 
To best our knowledge, we summary several typical state-of-the-art MAC protocols as examples in next section, regardless of the classifications, for wireless mesh networks with multi-beam antennas.

\section{Classic MAC Protocol Designs}
\label{Sec:claasic}
Wireless mesh networks using multi-beam smart antennas have received intensive attention recently due to its performance in higher gain and throughput. However, the ever popular contention-based MAC protocols, such as IEEE 802.11, are not too much effective for multi-beam antennas, and many challenging problems mentioned in section ~\ref{sec:challenges} need to be resolved. This section will present several classic MAC protocols to deal with these challenges.

In paper~\cite{07}, Wang et al. proposed a method, enhancing the performance of medium access control for WLANs with multi-beam access point. The authors proposed a novel MAC protocol to carefully address these challenging problems and improve the communication efficiency. Their design also considers the backward compatibility whereby an IEEE 802.11 terminal can transparently access a multi-beam access point.

\begin{figure}
  \centering
  \includegraphics[width=8cm]{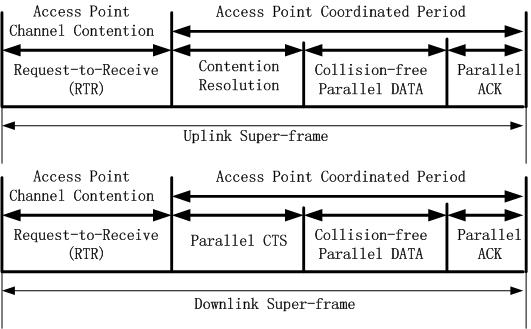}
  \caption{Uplink and Downlink Super-frame}
  \label{fig:0702}
\end{figure}

In~\cite{07}, the authors assume their system configured with multi-beam access points, omni-directional mobile nodes, and a single frequency channel, and they provided a distributed MAC layer solution based on IEEE 802.11 DCF. The data link of MAC protocol is separated into two types: Uplink medium access control and Downlink medium access control. The basic idea of their MAC protocol is to introduce a timing-structure to facilitate multiple handshakes (sequential or overlapping) before parallel collision-free data transmissions. Fig.~\ref{fig:0702} shows the outline both of uplink and downlink super-frame in their design.

For uplink medium access control, all the nodes having uplink packets in the queues contend for the channel access during the contention period. They design a contention resolution scheme to facilitate multiple nodes to win out, correspondingly, the winning nodes will be collision-free with each other. Thereby, the ``collision free" is actually a state that the winning nodes will not collide with each other when the winning nodes simultaneously send data to or receive ACK from the access point. Also, different winning nodes should be in different beam-sectors in order to be collision-free. In wireless mesh networks, the access point needs to contend for the channel as an ordinary node. In this case, the access point can send request-to-receive (RTR) messages in a higher priority over mobile-terminals to exchange RTS messages. For downlink medium access control, the access point in WMN needs to buffer the location information of each node. They provide two network scenarios: static and mobile. In the static scenario, the location information, obtained during the association process, is buffered in a static manner. While in the mobile scenario, the beam-location information of each node keeps cached by AP and it could be updated reactively whenever there is data needed to be exchanged or could be updated proactively using the periodic polling/probing initiated by the access point. 

In paper ~\cite{08}, Emad et al. proposed a distributed asynchronous directional-to-directional MAC protocol for wireless Ad hoc networks. The existing MAC protocols assume that the nodes can operate in both directional and omnidirectional modes, however, using both modes simultaneously could lead to the asymmetry-in-gain problem. The authors proposed a directional-to-directional (DtD) MAC protocol, where both sender and receiver operate in a directional-only mode, and they also derive the saturation throughput of ad hoc network using DtD MAC. The DtD MAC protocol is fully distributed, does not require synchronization, eliminates the asymmetry-in-gain problem as well as alleviates the effect of deafness and collisions.

DtD MAC protocol assumes that sending nodes cache the location information about their neighbor nodes, which this information is later used to determine the direction in which it should first try to send directional RTS (DRTS)messages. While the idle nodes, say potential receivers, continuously scan through their antenna sectors to emulate omnidirectional antennas, the potential receivers lock in the respective direction and response with a directional CTS (DCTS) message once they hear a DRTS intended for themselves. Each node estimates and caches the Angle-of-Arrival (AoA) of any messages it overhears to estimate the direction of the intended receiving node. Before sensing the medium, the sending node checks, using a machine learning mechanism, its AoA cache to determine the receiver's most likely direction. For the nodes that overhear ongoing communication accordingly, it sets their Directional Network-Allocation Vector (DNAV) to refrain from interrupting the ongoing communications in these directions.

In paper~\cite{03}, Bao et al. proposed a distributed receiver-oriented multiple access (ROMA) channel access scheduling protocols for ad hoc networks with directional antennas, each of which can form multiple beams and commence several simultaneous communication sessions. ROMA protocol determines the number of links for activation in every time slot using only two-hop topology information, which is unlike the random access schemes that use on-demand handshakes or signal scanning to resolve communication targets. This ROMA protocol significantly improves the network throughput, as well as the delay, can be achieved by exploiting the multi-beam forming capability of directional antennas in both transmission and reception. ROMA protocol adopts a Neighbor-aware Contention Resolution(NCR) to derive channel access schedules for a node. The contention to the shared resource is resolved in each context according to the priorities assigned to the entities based on the context number and their respective identifiers, and then select the highest priorities to access the common resource without conflicts.

Each link of ROMA protocol has a weight that reflects the data flow demand over the link in the ROAM network topology, and this weight is determined dynamically by the head of the link which monitors traffic demands or receives bandwidth requests from the upper-layer applications. Nodes and links are assigned priorities based on their identifiers and the current time slot. When the current time slot is \textit{t}, the priority of a node \textit{i} can be expressed by

$i.prio = Hash(i\oplus t)\oplus i$

where the sign $\oplus$ is designated to carry out the bit-wise concatenation operation on its operands and has lower order than other operations, function \textit{Hash()} is a fast pseudo-random number generator.

ROMA is a link-activation receiver-oriented multiple access protocol that exploits the multi-beam forming capability of multi-beam adaptive array antennas. Given the up-to-date information about the two-hop neighborhood of a node and link bandwidth allocations, ROMA decides whether a node \textit{i} is a receiver or a transmitter in a wireless mesh network, and which corresponding links can be activated for reception or transmission during the time slot \textit{t}. Before the actual link activation at the transmitters, ROMA protocol has to decide the active incoming links of each node in reception mode.

In paper~\cite{02}, Chou et al. proposed a MAC protocol, named M-HCCA (Multi-beam AP-assisted HCCA), which fully take advantage of multi-beam antennas equipped at the AP. This protocol does not only boost the overall capacity of WLAN, but also support QoS (Quality-of-Service) and power consumption for individual mobile stations. This MAC protocol has the following attractive features: (1) This MAC scheme is a polling-based protocol, hence it can innately conquer the problems induced by carrier sensing or directional signals. (2) M-HCCA can adaptively adjust the sector configuration, employing the deterministic tree-splitting algorithm as its reservation scheme, to quickly resolve contention/collision and to increase data transmission parallelism within the bounded reservation time, thus it achieves high real-time throughput. (3) M-HCCA adopts the mobile-assisted admission control technique, run-time admission control mechanism, such that the AP can admit as many new streams as possible during the reservation procedure while not violating QoS guarantees made to already-admitted streams, even in a multipath environment. (4) M-HCCA utilizes the beam-location-aware polling-based access scheme to reduce energy waste on collision and retransmissions as far as possible, meanwhile, it solves the beam-overlapping problem and back/side-lobe problem. (5) M-HCCA can effectively alleviate the miss-hit problem by offering a location updating mechanism to promptly renew the beam-location information of a non-responsive station.

\begin{figure}
  \centering
  \includegraphics[width=8cm]{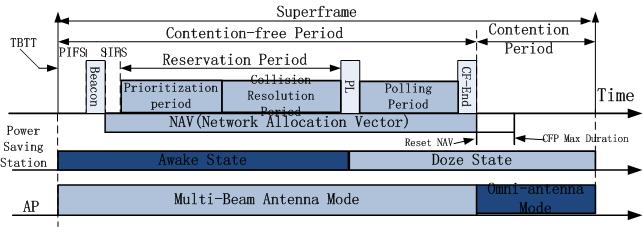}
  \caption{Super-frame structure for a WLAN with Multi-beam AP}
  \label{fig:0202}
\end{figure}

To get the Wifi services, a mobile station should first discover the presence of APs by passive scanning or active scanning. The AP normally operates in the multi-beam antenna mode during the Contention-Free-Period(CFP), except in a multipath rich environment. Fig.~\ref{fig:0202} shows the super-frame structure for a WLAN with multi-beam AP using M-HCCA protocol. If being equipped with the reconfigurable multi-beam antennas, the AP can adaptively adjust the sector configuration during the collision resolution period and the polling period to speed up the reservation process and minimize the average awake up time of polled stations. During the collision resolution procedure, M-HCCA utilizes an identifier-based tree-splitting algorithm to solve the collision problems. This algorithm is to use the stack to implement a pre-order traversal of the dimension splitting tree.

In paper ~\cite{05}, Vivek et al. proposed on-demand medium access in multi-hop wireless networks with multiple beam smart antennas. Traditional on-demand MAC protocols for omnidirectional and single beam directional antennas based on the DCF mechanism cannot take advantage of this unique capability of multiple beam antennas, as well as do not facilitate concurrent transmissions or receptions. This paper proposed a novel protocol, hybrid MAC (HMAC), which is backward compatible with IEEE 802.11 DCF. This protocol enables concurrent packet reception(CPR) and concurrent packet transmission(CPT) at multiple beamforming antennae. 

In the HMAC protocol, the design considers a wide-azimuth switched-beam smart antenna comprised of a multiple beam antennas array(MBAA), which is capable of calculating the exact angle of arrival of an incoming signal. HMAC is a cross-layer protocol that uses information from both the network and the physical layer for its operation. The novel features of HMAC include its channel access mechanism, an algorithm for mitigating deafness and contention resolution, jump back-off and role priority switching mechanisms for enhancing throughput. In HMAC, each node maintains its neighbor's information into the Hybrid Network Allocation Vector (HNAV) table, as well as a beam table (BT), which stores the backoff duration and two boolean variables for each beam. To eliminate the mitigating deafness problem, HMAC employs a hybrid approach based on the algorithm for mitigating deafness (AMD). To solve the channel contention problem, HMAC exploits a runtime sender-estimation algorithm (RSA). The node maintains a transmission probability for each neighbor node in its HNAV table, and the reciprocal of this probability is current estimate of the number of transmission in this beam. HMAC can also conduct the role priority switching, switching between transmitter and receiver roles, depending on the packets in its buffer. The rule is that as long as a data packet exists in the queue, the node gives priority to the transmission mode; otherwise, the reception mode supersedes.

In paper ~\cite{04}, Dhananjay et al. provided a performance evaluation of medium access control for multi-beam antenna node in a wireless LAN. In the paper, the authors used a wide-azimuth switched beam smart antenna system comprised of a multiple beam antenna array, and analyze and simulate the one-hop performance of CSMA as well as Slotted Aloha for these systems. The paper also investigated the problem of synchronization for multiple beams in CSMA.  These results show that, under heavy offered load conditions, CSMA is a good choice with nodes that multiple-beam smart antenna, despite the performance loss due to the beam synchronization. Also, CSMA protocol provides a stable throughput approaches unity and is invariant to fluctuations in the offered load. However, the performance of slotted aloha drastically reduces beyond optimum offered loads, although slotted Aloha is capable of higher peak throughput in a narrow range of offered loads as more switched beams are employed.

The analytical model of CSMA is basic CSMA, which is no handshake, no ACK. The carrier sense is performed by a node on the different channel than the one used for transmitting data, which the channel is comprised of a set of narrow-band tones. The model assumes that the estimated receiver beam number is mapped to a discrete tone, and the omnidirectional range of the tone must be set to reach all members that potential lie within a receiver beam, meanwhile it must keep as low as possible to reduce the probability of misclassification. The analytical model of slotted aloha is similar to that of CSMA, with some modification. They use the same concept, as within CSMA, of a regeneration cycle to estimate the throughput, whose cycle is a subchannel or a beam consisting of an idle period (no terminal has a packet to send), a busy period (one or more terminals have the packet for transmission).

In the paper ~\cite{08}, Karthikeyan et al. presented a medium access control in Ad Hoc networks with multiple inputs multiple outputs (MIMO) links. MIMO links can provide extremely high spectral efficiency in multipath channels by simultaneously transmitting multiple independent data streams in the same channel. The unique characteristics of MIMO links coupled with several key optimization considerations necessitate an entirely new MAC protocol. The authors present a centralized algorithm called stream-controlled medium access(SCMA) that has the key optimization considerations incorporated in its design. 

There exist some unique characteristics for the relevant physical layer. The incoming data is demultiplexed into M streams and each stream is transmitted out of a different antenna with equal power, at the same frequency, same modulation format, and in the same time slot. One of the aims of their work is to exploit the spatial multiplexing gain to increase the capacity of the system. In these designs, they use the centralized stream-controlled medium access protocol for ad hoc networks with MIMO links, which based on the observation about the receiver overheating problem: There exists a specific subset of links in the network that contributes to the lack of receiver overloading when performance pure stream control. To control the overflows of the bottleneck links, they use the schedules in the nonstream controlled fashion, such links can essentially be removed from further scheduling considerations, leaving the scheduling algorithm with only independent contention regions within which pure stream control can be employed. There is two scheduling component involves scheduling of bottleneck links in a nonstream controlled manner; scheduling of the nonbottleneck links in the network based on pure stream control.

In the paper ~\cite{10}, Dhananjay et al. proposed a novel MAC layer protocol for space division multiple access in wireless ad hoc networks. Most recent MAC protocols using directional antennas for wireless ad hoc networks are unable to attain substantial performance improvements because they do not enable the nodes to perform multiple simultaneous transmissions/receptions. The authors, in this paper, propose a MAC layer protocol that exploits space division multiple access thus using the property of directional reception to receive more than one packets from spatially separated transmitter nodes.

Space division multiple access employs time division duplex (TDD) between transmission and reception, thus this system does not require two separate antenna systems. It harnesses parallelism, at the same time, in the reception process, improving the throughput at a node, this needs those prospective transmitters to need to be synchronized to a receiver. Since each of their transmissions is dictated by a potential receiver, the transmitting nodes cannot synchronize their own transmissions to others. The system uses the receiver-initiated approach to achieve the time synchronization for receptions.

In the paper~\cite{09}, Yang-Seok et al. proposed a complementary beamforming. This paper proposed two new methods, called ``subspace complementary beamforming (SCBF)" and ``complementary superposition beamforming (CSBF)", to deal with the issue of complementary region, a region where some stations in the network cannot sense the directional signals (beams) often causing the hidden beam problem. The SCBF uses dummy data to ensure a controlled level of received energy in all directions of eigenvectors unused by downlink channels. Meanwhile, it enables CSBF which can also send data contained in the complementary beam. Using this method, the passive nodes in the network can receive ``broadcast" information, while the active nodes are engaged in the exchange of user-specific data. The effects of complementary beamforming can be achieved simply by increasing the transmit power of only one of antenna elements when space-division multiple access is not applied.

The main idea behind complementary beamforming (CBF) is that much less power is needed for unintended users to correctly detect the presence of transmission than that required for correct decoding of the transmitted packet in general. The SCBF technique creates a flat beam pattern in the ``otherwise hidden beam" directions, thus, the transmit power in all directions or locations is guaranteed to be great than a certain level. While CSBF applies a linear combination of the CBF vectors of SCBF. By modifying one of the SCBF downlink beamforming column vectors, its sidelobe level is increased without interfering with the desired beams. The overall objective of this design is to increase the probability that all the nodes can hear the channel activity so that they defer their transmissions. This means that the received power at any location in the service area should be greater than a certain threshold.

In the paper~\cite{Wang}, Wang et al. proposed a MAC protocol for multi-beam directional antennas, in which each beam-sector has its own control channel, and the communications among different beam-sectors are independent. It utilizes the directional network allocation vector (DNAV) to record the establishment processes. After sensing all the sectors of the multi-beam antenna, the protocol then uses the global assignment strategy to assign the directional communication channels. The whole protocol has several steps: connection initialization, channel contention, and data communication. These steps are sequentially in each beam sector.

In the papers~\cite{Biomo} and ~\cite{Biomo1}, Biomo et al. presented the case that full potential of Multi-Packet Transmission / Multi-Packet Reception (MPT/MPR)capability of multi-beam antennas can be unlocked to drastically reduce the end-to-end delay in ad hoc networks. The authors defined a formal optimization model for delay reduction and observed that the optimal end-to-end delay is attained when links are scheduled in the way that opportunities for MPT/MPR are maximized. Their results show that using the shortest routes, a widespread criterion in traditional routing protocols for ad hoc networks, results in higher delays, in which bridges among the routes incur waiting and rescheduling delay that adds to the end-to-end delay. 

In the paper~\cite{Grey}, Kuperman et al. presented a novel unslotted, uncoordinated ALOHA-like random access MAC policy for multi-beam directional systems that asymptotically achieves the capability of the network. In their setting, each communication node acts independently of one another. Its Multi-Beam Uncoordinated Random Access MAC (MB-URAM) does not make use of any reservation message, such as scheduling or RTS/CTS, and does not need to synchronize time slots or transmissions. The proposed protocol can asymptotically achieve the maximum possible throughput for any MAC approach, even a scheduled one. In the simulation, the authors considered practical considerations on the performance of MB-URAM, including power constraints, latency, beamwidth, and packet error rate.

In the paper~\cite{Fulvio}, Babich et al. discussed the design requirements for enabling multiple simultaneous peer-to-peer communication in IEEE 802.11 asynchronous networks in the presence of adaptive antenna arrays, and proposed two novel access schemes to realize multi-packet communication (MPC). One designed scheme, threshold access MPC (TAMPC) is based on a threshold on the load sustainable by the single-node; the other scheme, signal-to-interference ratio (SIR) access MPC (SAMPC), is based on an accurate estimation of the SIR and on the adoption of low-density parity check codes.  Both schemes rely on the information acquired by each node during the monitoring of the network activity, which is suitable for distributed and heterogeneous scenarios. Its setting considers the coexistence of legacy and non-legacy nodes equipped with different antenna systems.  Besides, both schemes are designed to be backward compatible with IEEE 802.11 standard, and their performances are compared to the theoretical one and to that of the IEEE 802.11n extension in a mobile environment.

In the paper~\cite{Brian}, Proulx et al. analyzed a new MAC protocol for multi-beam directional network via high-fidelity simulation using a real-time emulator. The work focuses on exploiting simultaneous transmission to create a distributed, low complexity, random access MAC protocol for multi-beam directional networks by exploiting the underlying physical abilities of a digital phased array (e.g., the ability to form receive beam a posteriori and form multiple transmit or receive beams). The protocol designed location tracking and power control methods to ensure the transmit beam is correctly pointed with the correct power.  Besides, the proposed scheme is able to track the state of a neighbor's random access protocol in order to drastically reduce the number of dropped packets and interference in the system. The proposed scheme was implemented on a new Extendable Mobile Ad-hoc Network Emulator (EMANE) model which allows for real-time, high fidelity performance evaluation.

In the paper~\cite{Wei}, Hong et al. reviewed multibeam antenna technologies for 5G wireless communications. The authors presented the key antenna technologies for supporting a high data transmission rate, an improved signal-to-interference-plus-noise ratio, and increased spectral and energy efficiency, and versatile beam shaping. Multi-beam antennas hold a great promise in serving as the critical infrastructure for enabling beamforming and massive MIMO that boost the 5G. The paper provided a thorough discussion on implementing multi-beam antennas on 5G settings.

\section{Discussions}
\label{Sec:dis}
We presented many state-of-the-art  MAC protocols for wireless mesh networks with multi-beam antenna in previous section. Here we discuss some issues that are highly related to MAC protocols .

\subsection{Problems with Current MAC Protocols}
\textit{1. Staleness of Beam Information}

As we had known, the beamforming information in wireless mesh network need to be obtained in advance and recorded in the Look-up Tables (LUT). However, due to the nodes mobility in wireless mesh networks, the staleness of beamforming information stored in LUT could occur if there exists the larger gap, than the beamwidth, between the cached and the actual beamforming information. The results obtained in paper~\cite{05} demonstrate that the performance of multiple beam antennas largely depends on the network topology. This problem could not be solved unless the beamforming information in the mesh topology is collected on a per-packet basis.

In case of handoff in WMN, when a node moves out from one subnet of a transmitter's beam to another subnet covered by another beam, all packets outbound for this node need to redirect to the new subnet. If the previous transmitter tries to transmit the packet to that moving node, the packet transmission addressed to this node fail. In this case, the packets addressed to this moving node in the current subnet need to be redirected to the new subnet. Also, it is important to detect these transmission failures at the MAC layer before being reported to the network layer. This handoff issue maybe launches an open area of research in wireless mesh networks.

\textit{2. Neighbor Nodes Discovery}  

To communication, the node should discover its neighboring nodes before data transmission. The neighbor discovery process does not lie within the domain of the MAC protocols, however, it has a great impact on the MAC layer's operation, especially in wireless mesh networks. In the WMN, the nodes should not only discover which nodes are within their communication range, but also identify the beamforming information of these neighbor nodes. The beamforming information is usually decided based on either the AoA estimation or the relative position of the nodes. However, traditional research usually relies on channel contention solution mechanisms, such as CSMA/CA in a distributed environment, to find the neighboring nodes. However, in current larger scales WMN, this issue will be very evident and complex or even falls into the deadlock to find their neighbor nodes.

The neighbor's address is provided by the network layer. The location-based beamforming usually requires additional hardware such as GPS and also implicitly assumes that a Line-of-Sight (LoS) exists between these nodes. This may not be accurate in multi-path and multi-hop environment considering a switched-beam antenna. It is better to use AoA estimation to find their neighboring nodes in WMN.

\textit{3. Multipath}

Multipath occurs, and very common in WMN, when multiple copies of the same signal are received by the receiver node from different directions in the multi-beam antenna. Multipath is primarily caused by reflection from terrestrial objects, such as buildings, and it thus is very high in urban areas. The multipath problem may result in a node activating several beams in the multi-beam antennas, thereby degrading the network performance. One possible solution can be a consideration that multiple beam smart antennas is installed on Mesh Routers (MRs) or Internet Gateways (IGWs) at the top of buildings, in this case, multipath may not affect the performance drastically.

\subsection{Prediction}
\textit{1. Heterogeneous Antennas}

In the previous sections, the analysis for most of the proposed MAC protocols is mostly based on homogeneous antennas which the antennas have the same characteristics. The considered antenna homogeneity includes the antenna type, the number of beams, radiation pattern or sometimes a beamforming reference direction. However, in large-scale WMN, there may exist different kinds of antennas, that is those communication nodes within the same WMN networks are equipped with heterogeneous antennas. It needs to carefully design the corresponding MAC protocols to deal with the heterogeneity-aware MAC protocols to fit large-scale wireless mesh networks.

\textit{2. Fairness}

One of the important characteristics of the MAC protocol is to provide fair channel access among the competing nodes in WMNs. It should not only focus on the optimization performance metrics, such as throughput and delay, but also consider the fairness among the competing nodes. One of the goals of MAC protocols is to enhance the spatial reuse, with continuous improvement of this goal, these MAC protocols usually result in the unfair medium access. Therefore, achieving fairness in wireless mesh networks with multi-beam antennas is a very challenging task that needs further exploitation.

\textit{3. QoS-aware Protocols}

With the pressing need of real-time services and running content-rich multimedia applications, Quality of Service (QoS) has become a vital component in wireless mesh networks. Most existing QoS-aware MAC protocols are limited to the single-hop wireless networks, however, although multi-hop wireless networks can improve the network performance at some extends, these concerns usually focus on the throughput and delay, little attention has been devoted to exploring this effectiveness in providing QoS guarantees, especially at the MAC layer. Thus, the researches need to focus more on the design of QoS-aware MAC protocols, as well as both intra-node and inter-node scheduling, in wireless mesh networks.

\section{Conclusion}
\label{Sec:Concl}
In this paper, we presented a comprehensive survey of MAC protocols in the wireless mesh network using multi-beam antennas. Theoretically, the capacity of WLAN can be considerably boosted by the use of multi-beam smart antennas. However, if the designers directly apply IEEE 802.11 to a WLAN with multi-beam antennas, it will inevitably encounter many challenges. The existing solutions to these challenges are based on DCF and hence are not suitable for multi-media applications. The design principles of MAC protocols need to exploit the benefits of multi-beam antennas and overcome the beamforming-related challenges.  Based on these aims, we enlisted and discussed the main challenges facing MAC protocols in wireless mesh networks with multi-beam antennas. Besides, we introduced the basics of multi-beam antennas, including multi-beam smart antennas and traditional medium access control protocols. Finally, we analyzed several classic MAC protocols for wireless mesh network using multi-beam antennas.

\end{document}